\documentclass[11pt, a4paper]{article}
\usepackage{hyperref}
\usepackage{graphicx}
\usepackage[hscale=0.75,vscale=0.8]{geometry}
\usepackage{amssymb, mathrsfs}

\begin{document}

\title{Remarks on matter-gravity entanglement, entropy, information loss and events\footnote{Contribution to the proceedings of the conference ``Progress and Visions in Quantum Theory in View of Gravity'', Leipzig, October 2018.   It is an extended written version of part of a talk which was given at the workshop ``The Mysterious Universe'', Mainz, Germany, 4 June 2019}}

\author{Bernard S. Kay$^*$ \medskip \\  {\small \emph{Department of Mathematics, University of York, York YO10 5DD, UK}} \smallskip \\ 
\small{$^*${\tt bernard.kay@york.ac.uk}}}

\date{}

\maketitle

\begin{abstract} 
I recall my \textit{matter-gravity entanglement hypothesis} and briefly review the evidence for it, based partly on its seeming ability to resolve a number of puzzles related to quantum black holes including the black hole information loss puzzle.   I point out that, according to this hypothesis, there is a quantity, i.e.\ the universe's \textit{matter-gravity entanglement entropy} -- which deserves to be considered the `entropy of the universe' and which, with suitable initial conditions, will plausibly increase monotonically with cosmological time.   In the last section, which is more tentative and raises a number of further puzzles and open questions, I discuss the prospects for a notion of `events' which `happen' whose statistical properties are described by this entropy of the universe.   It is hoped that such a theory of events may be a step on the way towards explaining how initial quantum fluctuations convert themselves into inhomogeneities in a seemingly classical universe. 
\end{abstract}

\section{Introduction}
\label{intro}

One of the big successes of inflation is that it explains the inhomogeneities in the CMB (the cosmic microwave background) and in the distribution of galaxies in terms of the quantum fluctuations in the initial quantum state of the universe.

\medskip

But how do \textit{quantum fluctuations} convert themselves into \textit{statistical inhomogeneities in a seemingly classical universe}?

\medskip

This seems to be a mystery.

\medskip

I'd like to talk about some ideas that were partly motivated by the hope to make progress on resolving this mystery. In the first nine sections, I recall my \textit{matter-gravity entanglement hypothesis} and briefly review the evidence for it, based on its seeming ability to resolve a number of puzzles related to quantum black holes including the black hole information loss puzzle.    As I will point out, according to this hypothesis, there is a quantity, i.e.\ the universe's \textit{matter-gravity entanglement entropy} -- which deserves to be considered the `entropy of the universe' and which, with suitable initial conditions, will plausibly increase monotonically with cosmological time.   In the last section, which is more tentative and raises a number of further problems and open questions, I discuss the prospects for a notion of `events' which `happen' -- whose statistical properties are described by this entropy of the universe.   It is hoped that such a theory of events will be a step on the way towards resolving the above mystery. 

\section{A classic thought experiment in nonrelativistic quantum mechanics}  
\label{classic}

\begin{figure}[h]
\centering
\includegraphics[scale = 0.4, trim = 7cm 18cm 2cm 2cm, clip]{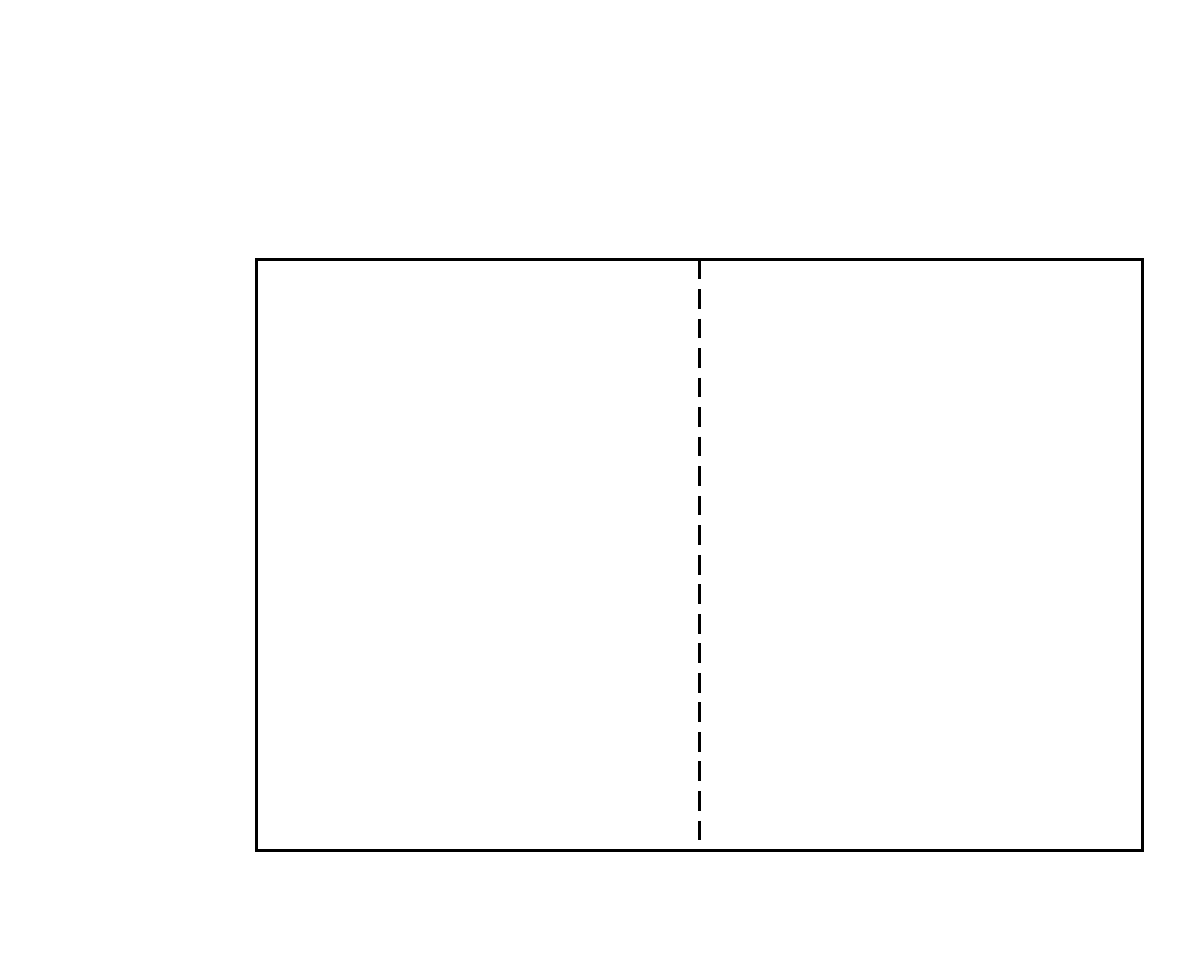}
\caption{\small{A box with a removable partition.}}
\label{fig:1}      
\end{figure}

Let's start with a more down-to-earth, but related, problem: a nonrelativistic quantum particle in a box in an initial pure (i.e.\ vector) state described by a Schr\"odinger wave function, $\psi$, initially confined to half of the box by means of a (removable) partition (Figure \ref{fig:1}).

If we remove the partition, then according to the unitary time evolution rule of quantum mechanics, the wave function will quickly evolve so as to fill the whole box and also, e.g.\ if it was in a stationary state (say the ground state) before the partition was removed, it will acquire a complicated shape and time-dependence.   But according to the unitary time-evolution rule, it will remain pure, and its von Neumann entropy, $S_{\mathrm{vN}}(|\psi\rangle\langle\psi|)$, will remain zero.\footnote{We recall that the von Neumann entropy of a density operator, $\rho$ is defined to be $-\mathrm{tr}\rho\ln\rho$.}  According to the standard rules \cite{vonN} of quantum mechanics, only upon an act of measurement\footnote{Here, we have in mind making a measurement but not recording the result.  See the second paragraph of Section (\ref{entropy}) for further discussion of this point.} by an observer, say of the observable which answers the question: \textit{Is the particle in the left half or the right half of the box?}, when the unitary time evolution rule gets suspended and the projection postulate applies, will $|\psi\rangle\langle\psi|$ change to an impure density operator, schematically, say 
\begin{equation}
\label{LR}
\rho = \frac{1}{2}|L\rangle\langle L| + \frac{1}{2}|R\rangle\langle R|
\end{equation}
and the von Neumann entropy jump from zero to $k\log 2$.

\section{Environment decoherence: pros and cons} 

We would like \cite{BellAg} to be able to arrive at a similar conclusion without invoking observers and measurements.    One way to do that is with the Environment approach to decoherence \cite{Zurek}.  This assumes that the unitary time evolution rule never gets suspended but needs to be applied at the level of a total closed system consisting of our system of interest (viewed now as an `open system') in interaction with its environment.  The wave function, $\Psi_{\mathrm{total}}$, of the total system consisting of our particle in our box together with its environment would be expected to be entangled between this system and its environment, and, soon after the partition gets removed, we would expect its entanglement entropy to increase by $k\log 2$.\footnote{Of course we would expect the particle in the box to always be entangled with the environment both before and after the removal of the partition -- albeit a little bit less entangled before its removal.   One could, if one wished to, make a mathematical model involving say a two-state system coupled to an environment where the total state of system and environment is unentangled (and hence the `state of the system' pure) before the removal of an idealized partition and entangled only afterwards.}

\begin{figure}[h]
\centering
\includegraphics[scale = 0.4, trim = 7cm 18cm 2cm 2cm, clip]{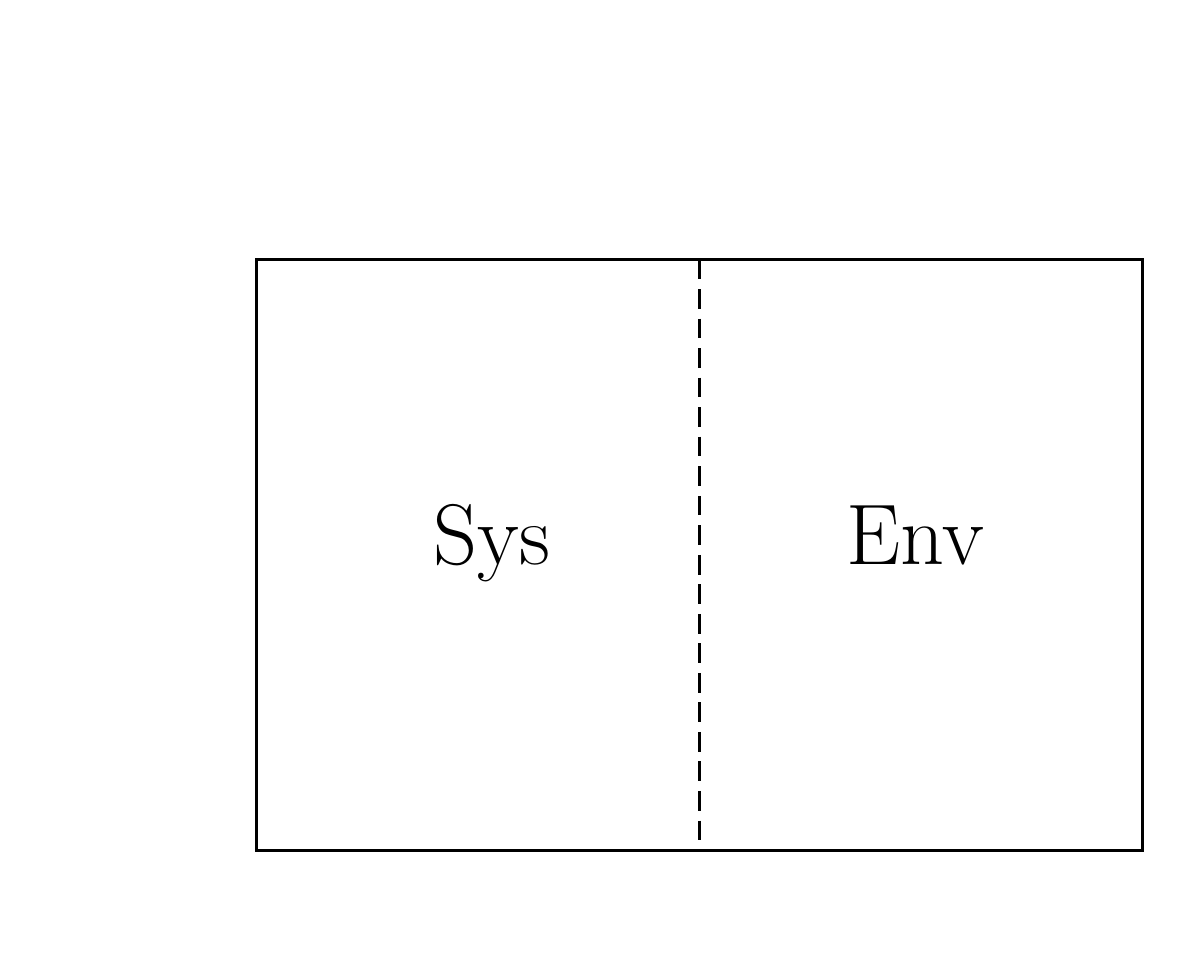}
\caption{\small{A schematic picture illustrating the traditional environment decoherence paradigm, with an open `system' in interaction with an `environment' and a partition indicating how the division of the total system into `system' and `environment' is changeable.  {\bf Note:} This picture is not to be confused with Figure 1, although the situation illustrated by Figure 1 is one possible candidate for `sys'.}}
\label{fig:2}      
\end{figure}

One problem, though, is that, in cosmology, there is (presumably) no environment that we can invoke, external to our universe.    We could, instead, take some subsystem of the universe and call it our `system' and call the remainder our `environment'.  But it's then a problem that there are many ways to do that.   Schematically, the boundary between system and environment in Figure \ref{fig:2} is slideable.

\section{Mathematical Interlude}
\label{interlude}

To spell out the basic assumptions here of what we mean by the environment approach to decoherence, they are that there is a total Hilbert space, ${\cal H}_{\mathrm{total}}$ which arises as the tensor product, ${\cal H}_{\mathrm{sys}}\otimes{\cal H}_{\mathrm{env}}$, of a \textit{system} Hilbert space, ${\cal H}_{\mathrm{sys}}$, and an \textit{environment} Hilbert space, ${\cal H}_{\mathrm{env}}$; and the density operator, $\rho_{\mathrm{total}}$, of the total system is, for all times, of the pure form $|\Psi(t)\rangle\langle\Psi(t)|$ where the vector $\Psi(t)\in{\cal H}_{\mathrm{total}}$ evolves in time according to a Schr\"odinger picture unitary time-evolution, i.e.\ such that $\Psi(t) = U(t)\Psi(0)$ where $U$ takes the form $U(t) = e^{-iHt}$ for a Hamiltonian, $H$, which arises as a sum of a matter Hamiltonian, a gravity Hamiltonian and an interaction term.

The density operator of the system, $\rho_{\mathrm{sys}}$ on the system Hilbert space, ${\cal H}_{\mathrm{sys}}$, is then defined to be the reduced density operator of $\rho_{\mathrm{total}}$ on ${\cal H}_{\mathrm{sys}}$ -- i.e.\ the partial trace of ${\rho}_{\mathrm{total}}$ over ${\cal H}_{\mathrm{env}}$.

The system von Neumann entropy, $S_\mathrm{vN}(\rho_{\mathrm{sys}})$, is then what we would normally think of as the entropy of the system due to its entanglement with the environment.

Notice that, by reversing the subscripts `sys' and `env' in everything, we can also define $\rho_{\mathrm{env}}$ and hence $S_\mathrm{vN}(\rho_{\mathrm{env}})$ -- or the entropy of the environment due to its entanglement with the system.   By a well-known result, thanks to our assumption that the total density operator is pure, 
the two entropies, $S_\mathrm{vN}(\rho_{\mathrm{sys}})$, $S_\mathrm{vN}(\rho_{\mathrm{env}})$ are equal and this is why we were able to refer, simply, to `the entanglement entropy' (i.e.\ between system and environment) in the previous paragraph and why (as we saw in the previous paragraph) that is the same thing as the physical entropy of the system.

The proof of this well-known result is an easy consequence of the \textit{Schmidt decomposition} theorem applied to $\Psi(t)$, and, since we will anyway need to refer to that theorem in the last section, we recall the statement of that theorem and the proof of our result here.   This states that for any vector, $\Psi$ in the tensor product ${\cal H}_A\otimes {\cal H}_B$, of two Hilbert spaces, there is an orthonormal basis, $\lbrace \alpha_a| a = 1, 2, 3, \dots\rbrace$ for ${\cal H}_A$ and an orthonormal basis, $\lbrace \beta_a| a = 1, 2, 3, \dots\rbrace$ for ${\cal H}_B$ and positive numbers, $c_a, \, a = 1, 2, 3, \dots$ such that
\begin{equation}
\label{Schmidt}
\Psi = \sum_{a=1}^{\infty}c_a \alpha_a\otimes \beta_a.
\end{equation}
Using this theorem, one sees that the partial trace of $|\Psi\rangle\langle\Psi|$ over ${\cal H}_B$ is $\sum_{a=1}^\infty c_a^2|\alpha_a\rangle\langle\alpha_a|$ and the partial trace of $|\Psi\rangle\langle\Psi|$  over ${\cal H}_A$ is  $\sum_{a=1}^\infty c_a^2|\beta_a\rangle\langle\beta_a|$ and hence the von Neumann entropy of both of these partial traces is 
$-\sum_{a=1}^\infty c_a^2\log(c_a^2)$.

\section{The matter-gravity entanglement hypothesis}

My hypothesis (see \cite{Kay98, KayNewt, KayAbyeeee,  KayEntropy, KayMatGrav}) which I want to tell you about, is that, at least at energies a bit below the Planck energy, there is an (effective) theory of quantum gravity which is a conventional quantum theory in which there's a total Hilbert space, ${\cal H}_{\mathrm{total}}$ which arises as the tensor product of a matter Hilbert space, ${\cal H}_{\mathrm{matter}}$ and a gravity Hilbert space, ${\cal H}_{\mathrm{matter}}$; and the total density operator of the total system is, for all times, of the form $|\Psi(t)\rangle\langle\Psi(t)|$ where the vector $\Psi(t)$ in the total Hilbert space evolves in time according to a Schr\"odinger picture unitary time-evolution for a Hamiltonian which arises as a sum of a matter Hamiltonian, a gravity Hamiltonian and an interaction term.   This is obviously mathematically entirely the same structure we had in the Environment paradigm with the substitution `matter' for `system' and `gravity' for `environment' and we further postulate that only operators on the matter Hilbert space are directly observable.   In short, we propose that there is a preferred system-gravity split according to which our system is matter and our environment is gravity.

\begin{figure}[h]
\centering
\includegraphics[scale = 0.4, trim = 7cm 18cm 2cm 2cm, clip]{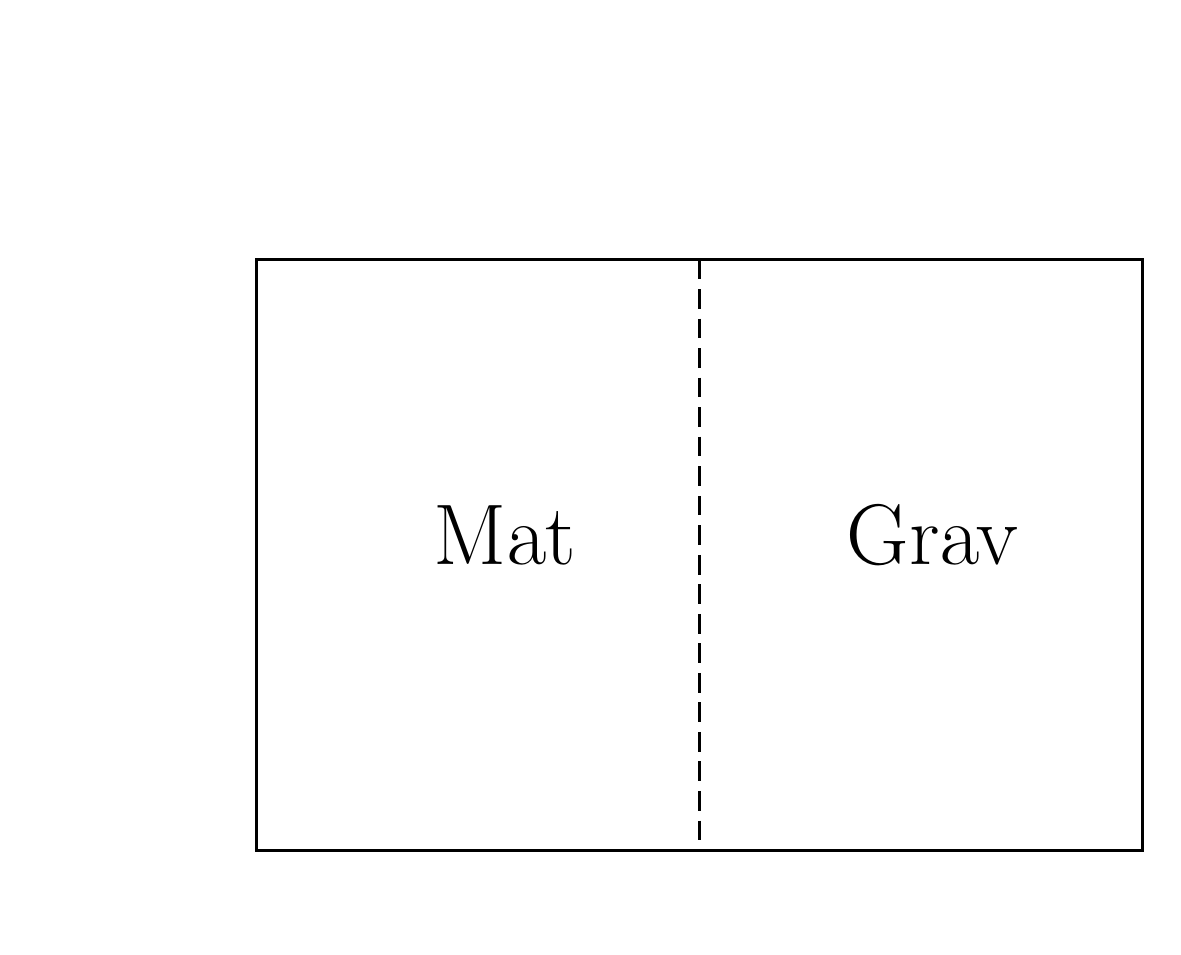}
\caption{\small{A schematic picture illustrating the matter-gravity entanglement hypothesis, according to which it is important to include the interaction of the matter sector of any closed system with gravity, with `matter' playing the role of `system' and `gravity' of `environment'.}}
\label{fig:3}      
\end{figure}

By the results in our Mathematical Interlude section above, we will then have
\begin{equation}
\label{equalentropies}
S_{\mathrm{vN}}(\rho_{\mathrm{matter}}) = S_{\mathrm{vN}}(\rho_{\mathrm{gravity}}) (= \hbox{the total system's `\textit{matter-gravity entanglement entropy}'})
\end{equation}
where $\rho_{\mathrm{matter}}$ is the reduced density operator of the matter and $\rho_{\mathrm{gravity}}$ is the reduced density operator of the gravity.

A further important part of our hypothesis is that these three equal quantities should be identified with the physical entropy, $S_{\mathrm{physical}}$ of our total closed system.

\section{The thermal atmosphere puzzle and its resolution} 

\begin{figure}[h]
\centering
\includegraphics[scale = 0.7, trim = 6cm 19cm 6cm 5cm, clip]{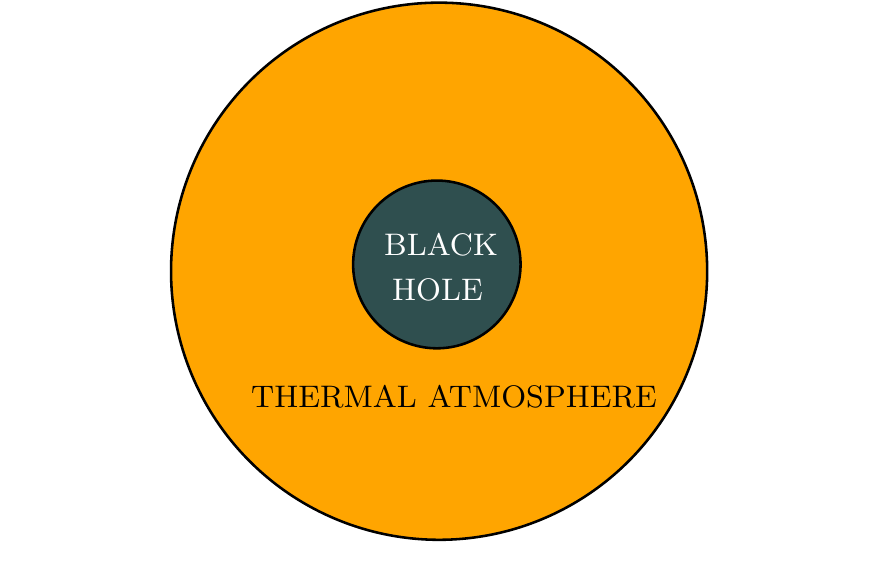}
\caption{\small{A schematic picture of a black hole in equilibrium with its thermal atmosphere in a box.}}
\label{fig:4}      
\end{figure}

The `\textit{thermal atmosphere puzzle}'  concerns a model closed system consisting of a black hole in equilibrium with its thermal atmosphere in a box (Figure \ref{fig:4}).   

The puzzle is \cite{Page, Wald} that there are ways of computing the entropy of this model system (obtaining approximately the Hawking value:  one quarter of the area of the event horizon) by calculating just the entropy of the black hole ($\approx$ the entropy of the gravitational field) and ways of computing its entropy by calculating the entropy of the thermal atmosphere ($\approx$ the entropy of the matter) and also, if one were to add these quantities, one would get twice the correct answer.  (For further discussion, including a discussion of how one can understand the calculation of black hole entropy in string theory on my matter-gravity entanglement hypothesis, see \cite{KayEntropy} or \cite{KayMatGrav} and references therein.)   

Clearly our equations (\ref{equalentropies}) seem to offer a neat resolution to this puzzle and indeed it was consideration of the thermal atmosphere puzzle that led me to propose the matter-gravity entanglement hypothesis.

\section{Resolution of the Black Hole Information Loss Puzzle -- and of the Second Law Puzzle}
\label{Resolution}

\begin{figure}[h]
\centering
\includegraphics[trim = 6cm 21cm 6cm 4cm, clip]{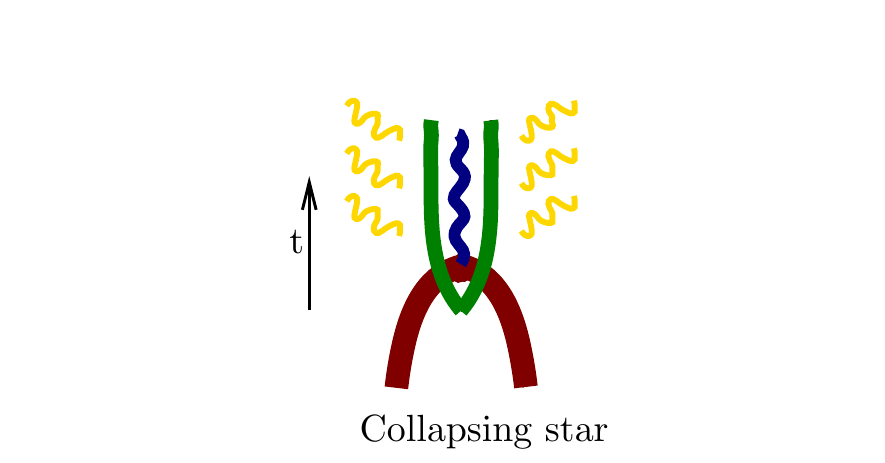}
\caption{\small{A schematic picture of the spacetime of a star which collapses to a black hole and then Hawking-evaporates.   The thick brown lines represent the boundary of the surface of a collapsing star, the green lines the horizon, the blue wiggly line the future spacetime singularity.  The thin yellow wiggles indicate the Hawking radiation.}}
\label{fig:5}      
\end{figure}

Turning to a model closed system consisting of a collapsing star in an otherwise empty, asymptotically flat universe -- which then collapses to a black hole and then Hawking radiates and eventually evaporates (Figure \ref{fig:5}), I argue that the above proposed new understanding of black hole entropy resolves the black hole information loss puzzle.   This puzzle was thought to arise from the fact that, during this process, the physical entropy of the total system is expected to always increase.   (And hence, if we think of information as negative entropy, information is expected to be lost.)

As far as I understand, the reason people thought there was a \textit{black hole information loss puzzle} is because they made the mistake of identifying the physical entropy of the total system with its von Neumann entropy -- the puzzle then being that since von Neumann entropy is a unitary invariant, it can't increase if we assume unitarity.   Once we, instead, identify physical entropy with \textit{matter-gravity entanglement entropy}, it can perfectly well increase under a unitary (Hamiltonian) dynamics.   In fact if we make the further assumption that \textit{the initial state, $\Psi(0)$, is unentangled (or at least has a low degree of entanglement) between matter and gravity}, we would expect it to get ever more entangled as time goes on and hence for its entropy (on our definition) to increase.

\medskip

Furthermore, a similar story would imply the entropy increase of any other closed system and, in particular, of the universe, if one assumes, likewise that its initial state has a low degree of matter-gravity entanglement, and thus offer an explanation for the second law.   So I would say that the black hole information loss puzzle is just a special case of the \textit{second-law puzzle}, and both are resolved by the matter-gravity entanglement hypothesis.

\section{Open Systems (Cups of Coffee)} 

\begin{figure}[h]6
\centering
\includegraphics[scale = 0.3, trim = 7cm 6cm 2cm 3cm, clip]{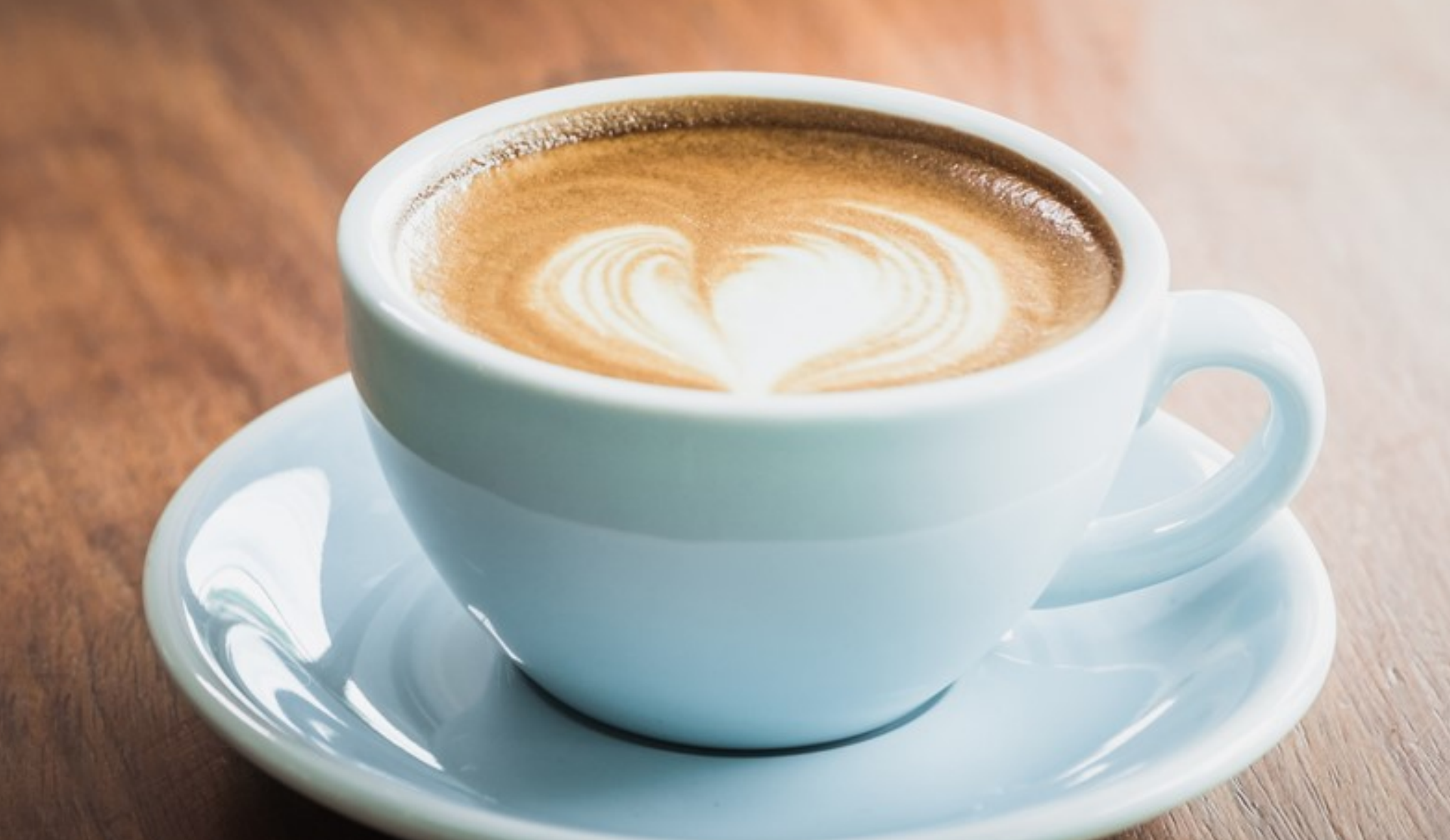}
\caption{\small{A cup of coffee}}
\label{fig:6}      
\end{figure}

Let me mention that, when I first proposed all this (back in 1998 \cite{Kay98}) people wrote to me and asked questions like:   \textit{What about the entropy of a gas in a box in the lab or of a cup of coffee (Figure \ref{fig:6})?  Surely it is crazy to think that could be due to entanglement with gravity?}  

But I think one can see that it's not actually far-fetched if one clarifies how one should think of open systems in my story.  The relevant picture (See \cite[Endnote (xii)]{KayAbyeeee} and \cite{KayEntropy} for further discussion) is now that shown in Figure 7.

\begin{figure}[h]
\centering
\includegraphics[scale = 0.4, trim = 7cm 13cm 2cm 2cm, clip]{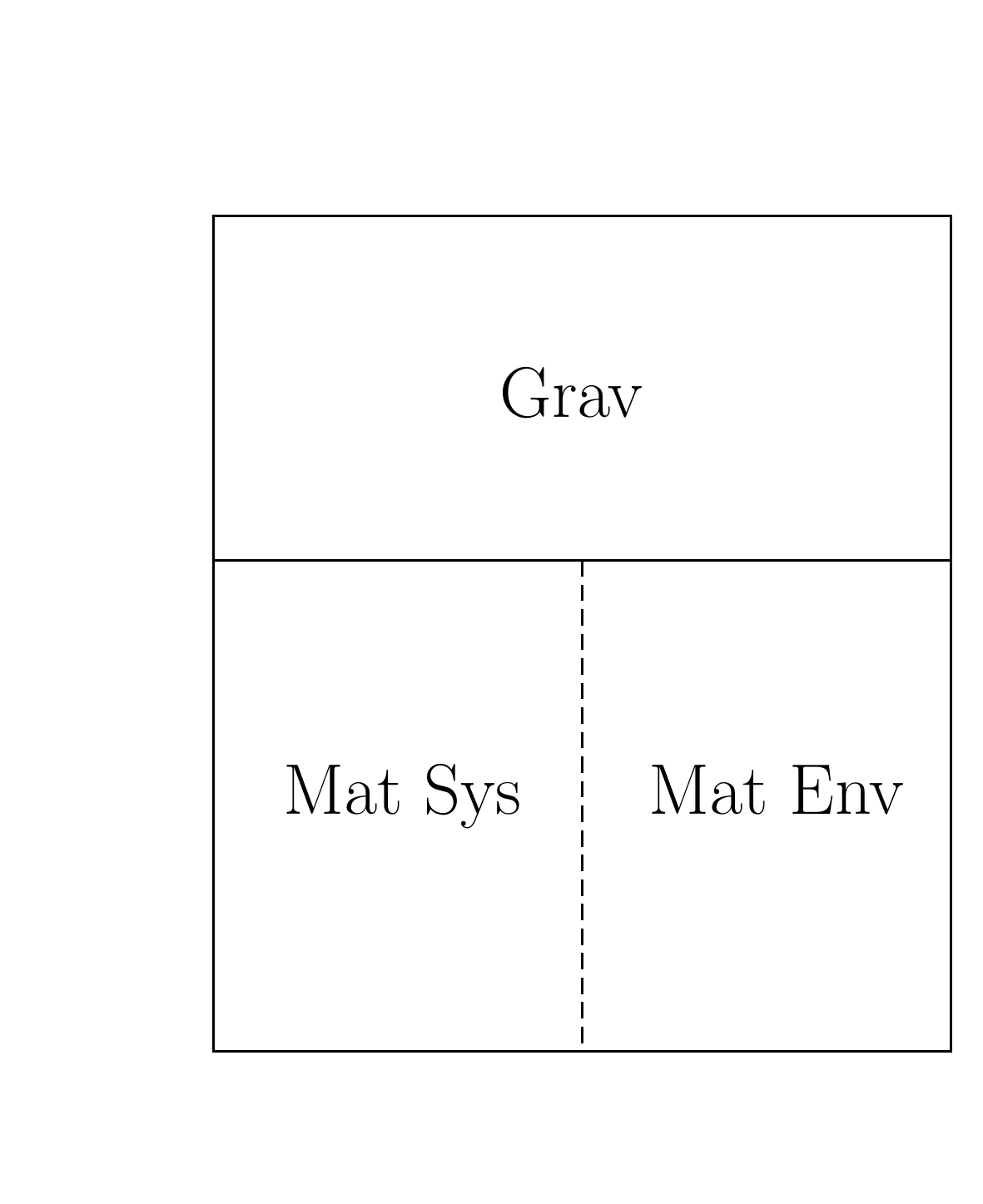}
\caption{\small{A schematic picture illustrating the matter-gravity entanglement paradigm as extended to apply to open systems, with an open `matter system' entangled with a `matter environment' as well as with the `gravity' sector which is a permanent part of the total environment.}}
\label{fig:7}      
\end{figure}

As this picture hopefully illustrates:   For a small open system (identified with a small subset of the matter degrees of freedom), the main environment is its matter environment.   Only for a large system (comparable in size to the whole universe) will the gravity environment start to dominate.  

\section{Events and time}  

\subsection{Unravelings}

A byproduct of our matter-gravity entanglement hypothesis is that, for a given, say unentangled (i.e.\ between matter and gravity) initial total density operator, a closed physical system has associated to it a well-defined time-evolving density operator for the matter degrees of freedom, $\rho_{\mathrm{matter}}(t)$ on ${\cal H}_{\mathrm{matter}}$ (whose von Neumann entropy at any time is the physical entropy of the total closed system).  Notice that, in the traditional Environment paradigm, there is no comparable preferred time-evolving density operator because of the slideability of the split between `system' and `environment'.  It is tempting (see \cite{Kay98, KayAbyeeee} and \cite[Section 3.3.2]{JuaSudKayV1}) to interpret the evolution of $\rho_{\mathrm{matter}}(t)$ as providing an objective replacement for the combination of unitary time evolution and occasional applications of the projection postulate of standard quantum mechanics (as e.g.\ formulated in \cite{vonN}) and thereby as going some way towards offering a solution to what is known as the `measurement problem' of quantum theory without the need to refer to observers and measurements -- as wished for, e.g., by Bell in \cite{BellAg}.   Indeed one might hope that $\rho_{\mathrm{matter}}(t)$ might play a similar role to the time-evolving density operator proposed in collapse models such as that of Ghirardi et al. \cite{GRW} known as GRW.  But as argued by Bell (in \cite{BellAg} and  \cite{BellJumps}) and also by Haag (see \cite{HaagEvents, HaagIndiv}) and others, one might hope that a fully satisfactory theory would not (or at least not only) be about a time-evolving density operator but (also) be about `events' which `happen'.

Here it is worth recalling that the prototype `GRW' collapse model \cite{GRW} for non-relativistic many-particle quantum mechanics was originally couched in terms of a certain master equation of general form:
\begin{equation}
\label{mastereq}
\dot\rho = L\rho
\end{equation}
for a time-evolving density operator, $\rho(t)$, where the linear operator, $L$, differs from the standard Schr\"odinger picture expression, $L_0=-i[H, \cdot]$,  by some particular extra terms involving certain new ad hoc constants of nature introduced in \cite{GRW}.  

We remark here, first, that, in view of our experience \cite{Zurek} with environmental decoherence models, we might expect our time-evolving density operator to approximately satisfy a master equation of the sort of (\ref{mastereq}), and in this sense to provide us with something like a collapse model -- but one where the various constants are not ad hoc new constants of nature but would be determined by the quantum gravitational Hamiltonian.   However, we would not expect $\rho_{\mathrm{matter}}(t)$ to \textit{exactly} satisfy such an equation.   (See again \cite[Section 3.3.2]{JuaSudKayV1} for more discussion.)

Secondly, in the paper ``\textit{Are there quantum jumps?}'' \cite{BellJumps}, Bell showed (based on the way that the authors of \cite{GRW} arrived at their time-evolution rule for their density operator) that the time-evolving GRW $\rho(t)$ can be interpreted as the density operator which describes the quantum statistics of a time-evolving quantum mechanical many-particle wave function $\psi(t)$ which, according to some stochastic process, every so often gets multiplied by a certain narrow function of the coordinates of one of the particles, centred on a certain random location called a `collapse centre', thus localizing the position of that particle and thereby leading to localization e.g.\ of pointers in measuring instruments.    As Bell argued in \cite{BellJumps}, one can then interpret the random occasions when such narrowings occur as the times at which `events' occur and the collapse centres as their locations.\footnote{It seems worth making the bibliographical remark here that many authors seem to inaccurately attribute what Bell did in \cite{BellJumps} to \cite{GRW} -- perhaps owing to the fact that, in \cite{BellJumps},  Bell himself appears to attribute what he does in \cite{BellJumps} to the authors of \cite{GRW}.}    However, it has since become clear that that interpretation corresponds to just one \textit{unraveling} out of many.  

Here we recall that an unraveling \cite{GisinBrunRigo} of a time-evolving density operator,
$t\mapsto\rho(t)$ on a Hilbert space, $\cal H$, is a set of (in general, not mutually orthogonal at each time, $t$) time-evolving (normalized) state vectors, $t\mapsto\psi_\mu(t)\in {\cal H}$ labelled by a parameter, $\mu$, which ranges over the elements of the sample space of some probability space, such that $\rho(t)$ is the mean of $|\psi_\mu(t)\rangle\langle\psi_\mu(t)|$.  For any time-evolving density operator, be it that of the GRW scheme, be it the $\rho_{\mathrm{matter}}(t)$ of interest here, there are many possible unravelings.   And, one expects (cf.\ \cite{GisinBrunRigo}) different unravelings of a given time-evolving density operator will lead to different interpretations in terms of events which happen, and, at least in the case of some unravelings, to different emergent classical descriptions.  (See also e.g.\ \cite{PerSahSud}.)

\subsection{A different possible approach to events which happen (with or without resets) and the symmetry puzzle}
\label{Different}

A seemingly different (i.e.\ from an unraveling) possible approach to associating a time-evolving density operator, $\rho(t)$ (be it the $\rho(t)$ of GRW or our $\rho_{\mathrm{matter}}(t)$ or any other time evolving density operator) with a notion of `events' which `happen' would be simply to declare that:

\medskip

\noindent
\textbf{Declaration}\quad\textit{For a given time-evolving density operator, $\rho(t)$, the set of possible events which can happen at a given time, $t$, is to be identified with the set of spectral projectors, $P_a$ (for $a$ for which $\lambda_a \ne 0$) in the spectral resolution, $\rho(t)=\sum_a\lambda_aP_a$, of $\rho(t)$ at that time $t$.  And the probability with which a given such event, $P_a$, occurs is then to be identified with $m_a\lambda_a$ where $\lambda_a$ is the eigenvalue of $\rho(t)$ belonging to the subspace on which $P_a$ projects and $m_a$ is its multiplicity (i.e.\ the dimension of the subspace).}

\medskip

It seems conceivable that (in the context of our matter-gravity entanglement hypothesis) such a notion of events might suffice, by itself, to account for quantum phenomena and this was the point of view implicitly adopted in \cite{KayAbyeeee}.    However,  below, we shall also discuss the possibility that the above Declaration (or the Modified Declaration below or the Alternative Declaration of the next subsection) needs to be supplemented by a \textit{reset rule}.\footnote{Most of this article, up to this point has reviewed earlier work of the author.  However, the remainder of this subsection and all of the next  subsections report on new research.  Also, as far as I am aware, at least some of what we say in the remainder of the paper is new.}  We shall first state such a rule appropriate to cases (such as collapse models such as GRW) where $\rho(t)$ satisfies a master equation of form (\ref{mastereq}).    For these cases, one can have a reset rule that prescribes that, at each of a certain set of (say, discrete) times, $t_1, t_2, \dots $ (following some initial time $t_0$ at which the initial state is assumed to be given) the master equation for $\rho(t)$ is suspended and one of the events, say the $a$th event, i.e.\ $P_a$,  that can happen at that time is selected, with probability $m_a\lambda_a$, and then the density operator at that time is \textit{reset} by replacing it by $\rho = m_a^{-1}P_a$ -- the subsequent time evolution until the next reset-time being, again, given by the master equation.
   
Our time-evolving density operator of interest, $\rho_{\mathrm{matter}}$, will of course not exactly satisfy a master equation of form (\ref{mastereq}), but one could still perhaps argue that the above reset rule might still be a viable rule to the extent that 
$\rho_{\mathrm{matter}}$ \textit{approximately} solves a master equation of that form.   But one could also criticise such an argument on the grounds that, even when it can be demonstrated that our time-evolving density operator approximately satisfies a master equation,  the demonstration of this will presumably rely on the assumption that there \textit{aren't} any resets.   And anyway one might well feel that an approximation of this sort has no business in a reset-rule!\footnote{One can understand the difficulty of doing without a master equation when resets occur as due to the fact that, to determine the subsequent evolution of $\rho_{\mathrm{matter}}(t)$ after some time at which it has been reset to one of its (normalized) spectral projectors, and in the easiest case in which the latter is one-dimensional, say $|\psi_a\rangle\langle\psi_a|$, we would expect $\rho_{\mathrm{matter}}(t)$ to be given (until the next reset-time) as the partial trace over gravity of the unitary time evolute, from the reset time to $t$, of the projector onto $\psi_a\otimes\gamma\in 
{\cal H}_{\mathrm{total}}$ for some vector, $\gamma\in{\cal H}_{\mathrm{gravity}}$.  But knowledge of $\psi_a$ is insufficient to determine what $\gamma$ has to be.   This difficulty is one of the motivations for the alternative notion of events of the next subsection for which no such difficulty occurs.}  

For the remainder of this subsection, we ask the reader either to suspend any worries about these unsatisfactory features or, alternatively, just assume we are discussing collapse models (such as GRW) in which $\rho(t)$ does exactly satisfy a master equation.   In the next subsection we shall consider an alternative notion of events which goes together with an alternative reset rule for 
$\rho_{\mathrm{matter}}$ which is free from these unsatisfactory features.  (But the experience of considering the not totally satisfactory reset rule in the present subsection should help to motivate what we do in the next subsection.) 

As for the discretisation of time, one might think of discretising, say with evenly spaced time-intervals, as just a mathematical device and that, ultimately, one should take some suitable continuous-time limit.  (As we shall remark in the next subsection, another point of view is possible for the different notion of events considered there.)  I understand \cite{FrohPriv} that Froehlich has considered a certain such sort of continuous-time limit of reset quantum states in the context of certain non-relativistic (and non-gravitational) quantum mechanical models involving a quantum mechanical system interacting with an environment which he takes to be forerunners to his proposal for resolving the measurement problem in \cite{Fro,FroRel}.   

Let me also mention here the recent work of Hollowood \cite{Hollo1} which, with somewhat different motivations from ours here, involves a formalism which, in some respects, resembles that discussed in the present section.   Also, an earlier paper of Hollowood, \cite{Hollo2}, applies a similar formalism to that of \cite{Hollo1} to a similar question to our opening question of Section \ref{intro} about the mystery of the formation of classical inhomogeneities in the universe -- albeit involving a `generic' system-environment split in place of our matter-gravity split here.
 
One can argue that resets as above are needed so as to be in accord with the assumption, made by many authors on the foundations of quantum mechanics, that something like such resets actually happens in nature.  See in this connection, e.g.\ where Landsman writes, in \cite{Lands}, 
``\textit{Decoherence turns superpositions into mixtures, i.e., it removes interference terms (if only approximately), but it gives no explanation why only one term in the ensuing mixture survives}.''  Such a view is of course also taken in many works on collapse models such as in Bell's version \cite{BellJumps} of \cite{GRW} which we mentioned above.   Let us also note that Haag in \cite{HaagIndiv} seems to take such a view when he refers to the need for an ``\textit{additional principle of random realization}''.  

Let us remark that if all the spectral projectors of $\rho(t)$ are one-dimensional for all times $t$, then restricting attention to the one-dimensional projector selected at each of the reset-times, $t_1, t_2, \dots$ (and up to an unimportant ambiguity in the phase of $\psi$ at each of those times) we will clearly obtain a discrete-time version of an unraveling\footnote{To spell out what is the discrete-time version of the unraveling:  Let us assume for simplicity that we restrict attention to a finite number of discrete reset-times, $t_1, \dots t_M$.   Then we may take the elements of the sample space to consist of M-tuples $(i_1, i_2, \dots , i_M)$, say, where $i_n\in \mathbb{N}$ is the number-label of the eigenvector of the time-evolute at time $t_n$ which happens at the reset at that time, on the assumption that the $i_1$th, $i_2$th, $\dots$, $i_{n-1}$th happened at the previous reset times -- it being understood that the relevant set of all eigenvectors at each reset time is numbered according to decreasing eigenvalue.  And the probability with which $\psi_{(i_1, i_2, \dots , i_M)}$ occurs will be the product of probabilities $\lambda_{i_1}\lambda_{i_1i_2}\dots\lambda_{i_1i_2\dots i_M}$ where $\lambda_{i_1\dots i_n}$ is the probability that the $i_n$th event will happen at reset-time $t_n$ given that, at the previous reset times, the $i_1$th, $i_2$th \dots $i_{n-1}$th events have happened.   (This could be illustrated by a tree-diagram such as Figure 3 in \cite{Fro}.)} -- albeit a non-symmetry violating unraveling, as we next discuss.

Whether or not the projectors are one-dimensional, and whether or not resets are assumed to happen, there is an important difference between the notion of events in the above declaration (with or without resets) and the events-interpretation of the unravelings considered in the Bell version \cite{BellJumps} of GRW and in many other collapse models.    Namely, if the time-evolving $\rho(t)$ were to possess a \textit{symmetry} (by which we shall always mean throughout this and the next subsection, a unitary [group of] operator[s] on the Hilbert space on which $\rho(t)$ acts [in the case of $\rho_{\mathrm{matter}}(t)$, that means on ${\cal H}_{\mathrm{matter}}$] that commutes with $\rho(t)$) then each of the spectral projectors in the above declaration would inherit that symmetry and also the symmetry would clearly be preserved by any resets -- whereas an individual trajectory belonging to a typical collapse-model unraveling would typically violate that symmetry.   This is a crucial point for the opening question of this article:  As remarked in \cite[Section 3.3.2]{JuaSudKayV1}, with a symmetry-non-violating unraveling, our matter-gravity entanglement hypothesis would not seem capable of explaining how cosmological structure could emerge that fails to respect the symmetry of an, assumed symmetric, initial state, whereas one would obviously want it to fail to respect it to explain inhomogeneities in our universe! We shall call this the \textit{symmetry problem}.\footnote{I am grateful to Daniel Sudarsky for pointing this symmetry problem out to me around 2010.}   Let us also mention that, in the context of inflationary scenarios, and in a framework based on a semiclassical version of the Einstein equations, in which a classical spacetime metric is coupled to the expectation value of a quantum stress-energy tensor,  Sudarsky and collaborators (see \cite{PerSahSud} and subsequent related papers) are pursuing a program to study the detailed implications for cosmological structure formation for a number of different relativistic versions of collapse models involving symmetry-violating unravelings. 

One possible way to overcome this symmetry problem, which would be effective when the spectrum of $\rho_{\mathrm{matter}}$ includes many spectral values which are degenerate, would be with the following Modified Declaration:

\medskip

\noindent
\textbf{Modified Declaration}\quad\textit{For a given time-evolving density operator, $\rho(t)$, the set of possible events which can happen at a given time, $t$, is to be identified with the set of one-dimensional spectral projectors, $|\psi_a\rangle\langle\psi_a|$, where $\{\psi_a|\ a = 1,2,3,\dots\}$ is an orthonormal basis of eigenvectors for $\rho(t)$.   Where there is degeneracy, causing the basis within each degeneracy subspace to be nonunique, it is to be chosen at random.}\footnote{Since the degeneracy subspaces of a density operator (for nonzero $\lambda_a$) are necessarily finite, one can define a random orthonormal basis by acting with a randomly chosen unitary (which is a meaningful notion for a finite dimensional Hilbert space) on an arbitrarily chosen orthonormal basis.   Similar considerations apply to the random choices in (\ref{SchmidtGrav}).}

The above notion of resets generalizes in an obvious and straightforward way to this modified declaration and it again only seems unproblematic in the case that $\rho(t)$ satisfies a master equation.   When it does, we obviously will again obtain a discrete-time version of an unraveling as discussed above for the case of no degeneracy -- but with the difference that the unraveling on the Modified Declaration may be symmetry-violating while the unraveling in the case of no degeneracy will not be.

\subsection{An alternative different possible approach to events which happen (again with or without resets)}
\label{Alternative} 

In this subsection we propose a possible alternative notion of events, and also of resets, suitable for our time-evolving density operator $\rho_{\mathrm{matter}}$.\footnote{Everything in this subsection would, of course, be relevant to any situation where there is a preferred system-environment split by replacing `system' by `matter' and `environment' by `gravity'.}  To motivate what that might be, let us first recall from Section (\ref{interlude}), that, underlying our 
$\rho_{\mathrm{matter}}$ there is a state-vector, $\Psi(t) \in{\cal H}_{\mathrm{total}}={\cal H}_{\mathrm{matter}}\otimes {\cal H}_{\mathrm{gravity}}$ which (in the absence of resets) evolves unitarily -- $\rho_{\mathrm{matter}}(t)$ being obtained from $\Psi(t)$ by taking the partial trace, $\mathrm{tr}_{{\cal H}_{\mathrm{gravity}}}(|\Psi(t_1)\rangle\langle\Psi(t_1)|)$, of $|\Psi(t_1)\rangle\langle\Psi(t_1)|$ over ${\cal H}_{\mathrm{gravity}}$.   Bearing this in mind, it seems natural to consider an event-notion where the events are state-vectors in the full matter-gravity (total) Hilbert space and are related directly to the underlying $\Psi(t)$ rather than to the $\rho_{\mathrm{matter}}(t)$ which is determined by it. 

\medskip

\noindent
\textbf{Alternative Declaration}\quad\textit{Given a unitarily time evolving state-vector, $\Psi(t)$ in 
${\cal H}_{\mathrm{total}}$, the set of possible events which can happen at a given time, $t$, is to be identified with the set of terms (after each is normalised to unit norm) in the Schmidt decomposition (\ref{Schmidt}) 
\begin{equation}
\label{SchmidtGrav}
\Psi = \sum_{a=1}^{\infty} c_a \psi_a\otimes \gamma_a
\end{equation} 
of $\Psi(t)$.  And the probability with which a given such event, $\psi_a\otimes\gamma_a$,  occurs is then to be identified with the square, $c_a^2$, of its coefficient in (\ref{Schmidt}).  Where more than one of the $c_a$ are equal, and, in consequence, the Schmidt decomposition is nonunique, the Schmidt decomposition is to be chosen at random.}

\medskip

As before, it is unclear whether or not this notion needs supplementing with a process of resets or not.  If we assume that it does need supplementing, then, given a $\Psi(t)\in {\cal H}_{\mathrm{total}}$  (which determines our $\rho_{\mathrm{matter}}$) we propose the following alternative notion of reset:

At each reset-time, $t_r$, one momentarily suspends the unitary time-evolution rule for $\Psi(t)$ and replaces $\Psi(t_r)$ by the single term $\alpha_a\otimes\gamma_a$  with probability $c_a^2$.   This will then evolve under the quantum gravitational unitary time-evolution rule until the next reset time.   We of course continue to define $\rho_{\mathrm{matter}}(t)$ at all times (including at reset times) as the partial trace of $|\Psi(t_1)\rangle\langle\Psi(t_1)|$ over ${\cal H}_{\mathrm{gravity}}$.

Concerning the status of the discrete times in this alternative notion of resets, we could again contemplate e.g.\ taking evenly spaced time-intervals and taking a continuous-time limit.  Or one might speculate, motivated by Penrose's ideas on the role of quantum gravity in state-vector reduction \cite{Penrose96}, that some yet-to-be-discovered more fundamental theory might be approximately described by supposing that a reset of $\Psi$ as considered above occurs whenever, in some sense still to be explored, the different gravitational states, $\gamma_a$, occurring in the Schmidt decomposition of $\Psi$, satisfy some criterion which roughly corresponds to the decomposition involving macroscopically distinct spacetimes.\footnote{Note that in \cite{KayNewt}, we pointed out that the matter-gravity entanglement hypothesis leads to some formulae which resemble those of Penrose.  Here we are speculating that we may need to \textit{combine} the matter-gravity entanglement hypothesis with Penrose's gravitational induced state-vector reduction ideas.}
 
One advantage of the above alternative notions of events and of resets over the notions discussed in the previous subsection is that we no longer need to rely on the existence of a master equation, which, at best is only approximately satisfied by $\rho_{\mathrm{matter}}(t)$.   Furthermore, there seem to be two ways in which our above alternative notion of events could possibly overcome the symmetry problem of the previous subsection.  First of all, it is possible that, say, an initial, say unentangled, total state, $\Psi=\psi\otimes \gamma$, for which the initial partial state of matter, $|\psi\rangle\langle\psi|$, has a certain symmetry, will (say, in the absence of resets) evolve into a total state at some later time whose partial trace over gravity violates that symmetry.  This would be expected if it was only the initial $\psi\in {\cal H}_{\mathrm{matter}}$ which respected the symmetry in question, while the symmetry did not extend to a symmetry on the total state $\psi\otimes\gamma\in {\cal H}_{\mathrm{total}}$.  Secondly, and even if the symmetry did extend in such a way, the symmetry could be violated at later times if the initial state in (\ref{SchmidtGrav}) had more than one of the $c_a$ equal.  For then the random-choice provision of the above `Alternative Declaration' would come into play and there's no reason why the random choices which actually get made would respect the symmetry.   This latter second way in which the symmetry problem could be overcome is rather similar to the way in which the symmetry problem got overcome with the `Modified Declaration' of Section (\ref{Different}).

\subsection{Entropy}
\label{entropy} 

If we adopt a stochastic events-based interpretation of our time-evolving 
$\rho_{\mathrm{matter}}(t)$ -- be it based on a choice of unraveling, be it based on a stochastic time-sequence of projectors as considered in Section (\ref{Different})\footnote{We shall postpone consideration of the alternative proposal of Section (\ref{Alternative}) to later in this subsection.} on the assumption that resets occur  --  then one might at first worry that one would lose the possibility of explaining the monotonic increase of entropy of a closed system which has an initially low entropy.   In the case of an unraveling, for all times, $t$, the state of the system may be assumed to \textit{actually be} the pure state $|\psi_\mu(t)\rangle\langle\psi_\mu(t)|$ for some particular sample-space element, $\mu$, and thus to have at all times entropy zero. In the case of the Declaration of Section (\ref{Different}) on the assumption that resets occur, whenever a reset occurs and $\rho_{\mathrm{matter}}$ gets replaced by ($m_a^{-1}$ times) one of its spectral projectors, its von Neumann entropy will decrease.   And, if that spectral projector happens to be a one-dimensional projector, say $|\psi_a\rangle\langle\psi_a|$ (as, of course, is always the case on the Modified Declaration of Section (\ref{Different})) the entropy would decrease to zero.   So it might seem that, rather than entropy increasing, it either remains zero for all times or increases for some times, but then, occasionally, decreases.

This is all of course closely related to what would happen, on the standard interpretation of quantum mechanics, in our particle-in-a-box thought experiment (see Section (\ref{classic})) when the observer performs a measurement to answer the question, \textit{Is the particle in the left half or the right half of the box?}.   Once the external observer updates their information by recording the outcome of their measurement, then (from that observer's point of view) the entropy will \textit{decrease} by $k\log 2$.

However, say in the case of a time-evolving density operator, $\rho(t)$, satisfying a master equation, $\dot\rho=L\rho$, and some unraveling thereof in terms of a time-evolving wave function, $\psi_\mu(t)$, where $\mu$ ranges over some appropriate sample space,  one can reconcile the fact that the von Neumann entropy of $\rho(t)$ may be nonzero and, say, increasing with time with the fact that, for a given $\mu$,  $\psi_\mu(t)$ is always a pure state, by realizing that $\rho(t)$ and $\psi_\mu(t)$ provide answers to two different sorts of question.   Namely $\psi_\mu(t)$ for some particular $\mu$ may be the right answer to the question:  \textit{What actually happens?}.   But, on the other hand, 
$\rho(t)$ may be interpreted as answering the question:  \textit{What are we able to predict given some assumption about the initial total state?}.  In fact it is reasonable to think of the von Neumann entropy of $\rho(t)$ as telling us something about the extent of unpredictability of the full set of all wave-function trajectories, $\psi_\mu(t)$, as the label, $\mu$, ranges over all the elements of the appropriate sample space.

This is reasonable because, say for the unraveling with discrete times discussed in Section (\ref{Different}), for some given time-evolving density operator, $\rho(t)$, with some fixed value $\rho(t_0)$ at some initial time $t_0$ and satisfying a given master equation $\dot\rho=L\rho$ at times other than reset times, then, after a reset at some reset-time, $t_i$, if one takes the \textit{statistical operator} -- by which we shall mean here the weighted sum of all the different possible time-evolving density operators that could occur, say between that reset-time and some time, $t$ (before the next reset time) -- weighted by the probabilities which they are likely to occur, then the result will coincide with the time-evolving density operator that would have obtained in the absence of a reset.   This follows simply from the linearity of the time-evolution operator (on density operators) $T=e^{L(t-t_i)}$.  In fact we have that the statistical operator as defined above equals  
$\sum_a\lambda_a T(|\psi_a(t_1)\rangle\langle\psi_a|(t_i))$ which equals $T\rho(t_i)$ which is equal to the density operator $\rho(t)$.   

We note, though, that, for the alternative notion of resets which we argued in Section (\ref{Alternative}) would be more appropriate for the time-evolving density operator $\rho_{\mathrm{matter}}$ (for which a master equation isn't expected to hold exactly) for a given initial total state vector, $\Psi=\sum_a c_a \psi_a\otimes \gamma_a\in {\cal H}_{\mathrm{total}}$ just before a reset-time $t_i$, the statistical operator subsequent to the reset at that time -- obtained by taking the appropriate corresponding weighted sum -- will differ from the density operator that would have obtained in the absence of a reset.  In fact, one easily sees that, if we denote by $U$ the unitary-time evolution operator on the total Hilbert space between times $t_i$ and $t$, then the former statistical operator will be $\sum_a c_a^2\ \mathrm{tr}_{{\cal H}_{\mathrm{gravity}}}\ (U|\psi_a\otimes\gamma_a\rangle\langle\psi_a\otimes\gamma_a|U^{-1})$, while the latter density operator will be\hfil\break
$\mathrm{tr}_{{\cal H}_{\mathrm{gravity}}}\ (U\sum_a c_a|\psi_a\otimes\gamma_a\rangle)(\sum_b c_b\langle\psi_b\otimes\gamma_b|U^{-1})$.   Presumably, to tell us something about the unpredictability of the notion of events of Section (\ref{Alternative}), the appropriate time-evolving entropy in the presence of resets is then the von Neumann entropy of the former statistical operator and not that of the latter density operator.   I.e.\ not the matter-gravity entanglement entropy of the time evolving density operator that would obtain in the absence of resets, albeit one might hope that, in many circumstances (for which resets are not too frequent) the two operators may closely approximate one another.

To attempt to sum up what all the above may have to tell us about the relevance of our matter-gravity entanglement hypothesis to our opening question in Section (\ref{intro}) about the mystery of the formation of classical inhomogeneities in the universe:  We have explored various ways in which our time-evolving density operator, $\rho_{\hbox{matter}}$ (or in the case of our Alternative Declaration of Section (\ref{Alternative}), our time-evolving vector, $\Psi\in {\cal H}_{\mathrm{total}}$, which underlies $\rho_{\hbox{matter}}$) can be interpreted in terms of `events' which `happen'.   Various issues arose:  Two related possible notions (described in the Declaration and Modified Declaration of Section (\ref{Different})) are problematic because $\rho_{\hbox{matter}}$ isn't expected to exactly satisfy a master equation.   Another possible notion (described in the Alternative Declaration of Section (\ref{Alternative})) seems to overcome some of those problems but, as we have seen, in the presence of resets, will have a statistical operator which differs from $\rho_{\hbox{matter}}$ (albeit this is not necessarily a problem).  Some possible notions of `events' which `happen' suffered from our symmetry problem.   But e.g.\ our Modified Declaration of Section (\ref{Different}) and our Alternative Declaration of Section (\ref{Alternative}) didn't suffer from that problem.   Clearly all we have discussed so far can only be regarded as tentative and it remains uncertain which (if any!) of the several approaches we have discussed will bear fruit.  And there anyway seems a considerable gap to be bridged between any of our notions of  `events' which `happen' and actual inhomogeneities in the actual universe.   Let us remark though that one might hope that this gap will be bridgeable by first establishing a clearer link between the present ideas (which are in the context of full quantum gravity) and the work in \cite{PerSahSud} and subsequent papers, which we mentioned earlier and which is based on the use of collapse models in the context of semiclassical gravity.  (See again also the discussion in \cite[Section 3.3.2]{JuaSudKayV1}.)

Another issue is that it is not clear whether a successful notion of `events' which `happen' requires resets to also happen in one of the senses we explained in Sections (\ref{Different}), (\ref{Alternative}) at all, or whether we can base a successful theory of events just on our (never reset) $\rho_{\hbox{matter}}$ (or on our never reset $\Psi\in {\cal H}_{\mathrm{total}}$).

Indeed it is perhaps worth reminding ourselves that it remains somewhat controversial whether a time-evolving density operator is insufficient for physics and whether we need a notion of `events' which `happen' at all.   And for good reasons:  If there are events which happen, we seem to have very little indication about how frequently, or (if the events under consideration can be somehow localized in space) how densely in space, they happen.  The fact that we have very little indication of this is of course closely related to the fact that there is considerable freedom, in the standard approach to quantum mechanics \cite{vonN}, as to where to place the von Neumann/Heisenberg cut between system and observer.  Relatedly, as far as I am aware, there is, as yet, no experimental evidence that favours (say) collapse models over other approaches to the measurement problem.  In fact (even though we may dislike it) it appears to remain difficult (at least as far as laboratory physics is concerned) to refute the view, as expressed, e.g.\ in \cite{BubPito}, that Bell, in \cite{BellAg}, led us astray when he urged us to formulate a theory without notions such as `observers' and `measurement' and `information'; that quantum mechanics is actually about updating our `information'; and that collapse models are misconceived! 

There would seem to be less reason, though, to worry that such controversies might impinge on our proposed resolutions of several puzzles in Section 1 to 9, for which the question of whether or not there are `events' which `happen', and whether or not there are resets, was seemingly not of great relevance and for which we relied mainly on the expected properties of the (von Neumann) entropy of our time evolving $\rho_{\hbox{matter}}$   (equivalently on the matter-gravity entanglement entropy of our $\Psi\in {\cal H}_{\mathrm{total}}$.)   Fortunately, as we saw earlier in this subsection, there are reasons to hope that the value of the entropy of a given closed system may not be too greatly affected by whether or not there are `events' which `happen' or whether or not there are resets and may therefore not depend very much on how any of the latter notions are defined.

\subsection{Time}
\label{Time} 

Aside from issues related to `events', another potential criticism of our matter-gravity entanglement hypothesis is that it seems to be tied to an overly na\"ive notion of `time' which seems at odds with the nature of time in general (or even in special) relativity; although in its favour is the fact that, in specific contexts, such as cosmology or our model black hole scenarios, there is a natural preferred time, with which it can be identified.

Here we content ourselves with noting  that the traditional interpretation of quantum mechanics (as formulated, say, in \cite{vonN}) seems capable of generalization at least to cope with special relativity and with quantum theory in a (fixed) curved spacetime (see e.g.\ \cite{FewVerPap, FewProc}).   Also proposals have been made e.g.\ by Bedingham \cite{Bedin} to generalize collapse models to relativistic theories in Minkowski space and (following different lines to those of \cite{Bedin}) by Sudarsky and collaborators (see e.g.\ \cite{JuaSudKayV1, JuaSudKay} and references therein) to curved spacetime.   We remark, concerning the curved spacetime formulation of GRW in \cite{JuaSudKayV1, JuaSudKay}, that collapses are argued to occur in local regions and affect what subsequently happens only in the future light-cone of those regions, in consistency with relativistic causality, even though collapses take place, \textit{a priori} on Cauchy surfaces.

Very recently, Froehlich has proposed a relativistic version \cite{FroRel} of his approach to quantum mechanics  which might be regarded as a sort of collapse model.   However, it is a rather different collapse model from the traditional GRW type and seems free from the sort of ad hoc assumptions and from the ad hoc introduction of new constants of nature in \cite{GRW, BellJumps} and in \cite{JuaSudKayV1, JuaSudKay}.   It is formulated in terms of algebraic quantum field theory (see e.g.\ \cite{FewRej}) and relies on the special features of the algebras for the interiors of future light cones in theories (say in Minkowski space) containing massless particles.   It replaces the ad hoc rules of collapse models by a mathematically natural criterion for when/where collapses occur/events happen.   It promises to share with \cite{FewVerPap, FewProc} and \cite{JuaSudKayV1, JuaSudKay} the feature that collapses occur locally and affect what subsequently happens only within future light cones.   The resulting theory seems to offer a theory of events, consistent with special relativity, along the lines desired by Haag.

Thus approaches to the measurement problem, and to providing quantum theory with a notion of `events' which `happen', do seem generalizable from a nonrelativistic context to the contexts of a fixed background flat or curved spacetime.
One might hope that such approaches may hold some clues for overcoming the presently seemingly overly na\"ive notion of `time' in the present formulation of our matter-gravity entanglement hypothesis (and thereby in the resolutions to the several puzzles related to quantum black holes and the Second Law which we discussed in Sections 1-9, which were based on that hypothesis).  Amongst the difficulties though is the fact that the latter hypothesis presupposes, as its context, a full theory of quantum gravity, where we expect e.g.\ that the notion of `light-cone' (which plays such an important role in those flat and curved spacetime approaches) can, at best, only be an approximate notion, subject itself to quantum fluctuations.   

\bigskip

\section*{Acknowledgements} 

I am grateful to Felix Finster, Domenico Giulini, Johannes Kleiner and 
J\"urgen Tolksdorf, organizers of the conference ``Progress and Visions in Quantum Theory in View of Gravity'', MPI, Leipzig, 1-5 October 2018 for inviting me as a discussant and inviting me to contribute an article to the proceedings.   This article is based on part of a talk, prepared and presented at the workshop ``The Mysterious Universe: Dark Matter -- Dark Energy -- Cosmic Magnetic Fields'' May 20 - June 7, 2019.  I am grateful to Daniel Wyler, Juerg Froehlich, Klaus Fredenhagen, Ruth Durrer and Pedro Schwaller for inviting me to this workshop and to the Mainz Institute for Theoretical Physics (MITP) of the DFG Cluster of Excellence PRISMA$^+$ (Project ID 39083149) for its hospitality and support.    At both meetings, I benefited from listening to many talks and from discussions related to the topics discussed here. I wish particularly to thank Juerg Froehlich for several very valuable discussions.  I also thank Roger Colbeck for a helpful discussion about the physical interpretation of time-evolving density operators.

\end{document}